\documentclass{article}
\usepackage{spconf,amssymb,amsmath,epsfig}
\usepackage{textcomp}
\usepackage{amsfonts,amsthm,amsopn}
\usepackage{bm}
\usepackage{hyperref}
\usepackage{url}
\usepackage[capitalise]{cleveref}
\usepackage{graphicx,caption,lipsum}
\usepackage{booktabs,multirow}
\usepackage{placeins}
\usepackage{algpseudocode,algorithm}
\usepackage{color}
\usepackage[table,xcdraw]{xcolor}
\usepackage{bbm}
\usepackage{enumerate}   
\usepackage{adjustbox}
\usepackage[normalem]{ulem}
\usepackage{fixltx2e}
\usepackage[toc,page]{appendix}
\usepackage{makecell}
\usepackage{soul}

\usepackage{color}
\definecolor{mypink1}{rgb}{0.858, 0.188, 0.478}
\newcommand{\hq}[1]{{\color{mypink1}{\emph{#1}}}}
\renewcommand{\hq}[1]{}

\title{Self-Supervised Speaker Verification with Simple Siamese Network and Self-Supervised Regularization}

\name{{Mufan Sang$^{1\ast}$, Haoqi Li$^{2}$, Fang Liu$^{2}$, Andrew O. Arnold$^{2}$, Li Wan$^{2}$}}
\address{$^{1}$The University of Texas at Dallas, TX, USA \\
$^{2}$Amazon AWS AI, USA \\
{\small \tt mufan.sang@utdallas.edu},
{\small \tt \{haoqili,amzfang,anarnld,lliwan\}@amazon.com}}

\graphicspath{{./image/}}

\begin{document}

\maketitle

\ninept

\begin{abstract}
Training speaker-discriminative and robust speaker verification systems without speaker labels is still challenging and worthwhile to explore. In this study, we propose an effective self-supervised learning framework and a novel regularization strategy to facilitate self-supervised speaker representation learning. Different from contrastive learning-based self-supervised learning methods, the proposed self-supervised regularization (SSReg) focuses exclusively on the similarity between the latent representations of positive data pairs. We also explore the effectiveness of alternative online data augmentation strategies on both the time domain and frequency domain. With our strong online data augmentation strategy, the proposed SSReg shows the potential of self-supervised learning without using negative pairs and it can significantly improve the performance of self-supervised speaker representation learning with a simple Siamese network architecture. Comprehensive experiments on the VoxCeleb datasets demonstrate that our proposed self-supervised approach obtains a 23.4\% relative improvement by adding the effective self-supervised regularization and outperforms other previous works.

\end{abstract}

%

\vspace{-2.5ex}
\section{Introduction}
\vspace{-1.0ex}
Speaker verification (SV) is a task to identify the true characteristics of the speaker from speech samples and to accept or discard the identity claimed by the speaker. Nowadays, automatic speaker verification technologies and systems have been widely used in many e-commerce applications, general business interactions, and even forensics. Additionally, speaker representations can also be crucial to speaker diarization, automatic speech recognition, speech synthesis, and voice conversion tasks. Although great progress has been made on supervised SV systems with various proposed deep neural network architectures~\cite{snyder2017deep, snyder2018x, desplanques2020ecapa}, novel loss functions~\cite{wan2018generalized, chung2020defence, wang2018cosface, deng2019arcface}, different pooling methods~\cite{cai2018exploring, okabe2018attentive}, and frameworks for domain mismatch~\cite{sang2020open, bhattacharya2019generative, sang2021deaan}, they inevitably rely on the availability of large-scale annotated datasets. However, manual annotation of large-scale datasets can be extremely costly and privacy-sensitive when involving biometrics. Thus, it is worthwhile to explore learning meaningful speaker representations by leveraging a large amount of unlabeled data.    

\renewcommand{\thefootnote}{\fnsymbol{footnote}}
\footnotetext[1]{Work performed while Mufan Sang was interning at Amazon AWS AI.}
Self-supervised learning (SSL) can utilize unlabeled data to learn useful representations, thereby it has received more and more attention in many fields recently. In the computer vision domain, contrastive learning-based methods like SimCLR~\cite{chen2020simple} and MoCo~\cite{he2020momentum} introduce positive and negative data pairs and encourage latent representations of positive pairs to be similar and those of negative pairs to be dissimilar. Although these approaches can explicitly avoid mode collapse in this way, both of them require a large number of negative samples to achieve decent performance. Recently, other approaches~\cite{grill2020bootstrap, chen2021exploring} using asymmetric architectures can achieve competitive performance without explicit strategies for avoiding collapse (e.g., using negative samples). While some of these methods are not well understood, the core idea behind these self-supervised learning methods is aiming to learn invariant representations from unlabeled data under different data augmentations. For speaker recognition, researchers have also explored contrastive learning-based self-supervised learning methods~\cite{inoue2020semi, huh2020augmentation, nagrani2020disentangled, chung2020seeing, xia2021self}. In~\cite{huh2020augmentation}, an augmentation adversarial training strategy is introduced to learn channel-invariant speaker representations. In~\cite{xia2021self}, it utilizes a prototypical memory bank to improve the performance of the MoCo speaker embedding system. However, these methods rely on complex training strategies (e.g., adversarial training), architectures (e.g., momentum encoder, large size dynamic queue), or a huge number of negative samples.  

In this study, we propose a novel self-supervised speaker representation learning method without involving adversarial training, momentum encoder, clustering for pseudo labels, or a large number of negative samples. Our framework consists of a self-supervised learning module and a novel self-supervised regularization (SSReg) module based on the Siamese network architecture. To the best of our knowledge, this is the first study to explore and propose a self-supervised regularization strategy for improving the performance of self-supervised speaker representation learning. Instead of using both positive and negative pairs as contrastive learning, the proposed SSReg can learn meaningful speaker embeddings using only positive sample pairs with a simple Siamese network. It pays attention to the similarity between latent representations of randomly segmented positive pairs exclusively, without using negative pairs. Based on the proposed strong online data augmentation strategy, experimental results indicate that the proposed SSReg can significantly boost the performance of self-supervised speaker verification. Our method outperforms many previous works with a simpler and more effective architecture.  

\begin{figure*}[th]
\centering
\scalebox{0.91}
{
\includegraphics[width=14cm,height=5.3cm]{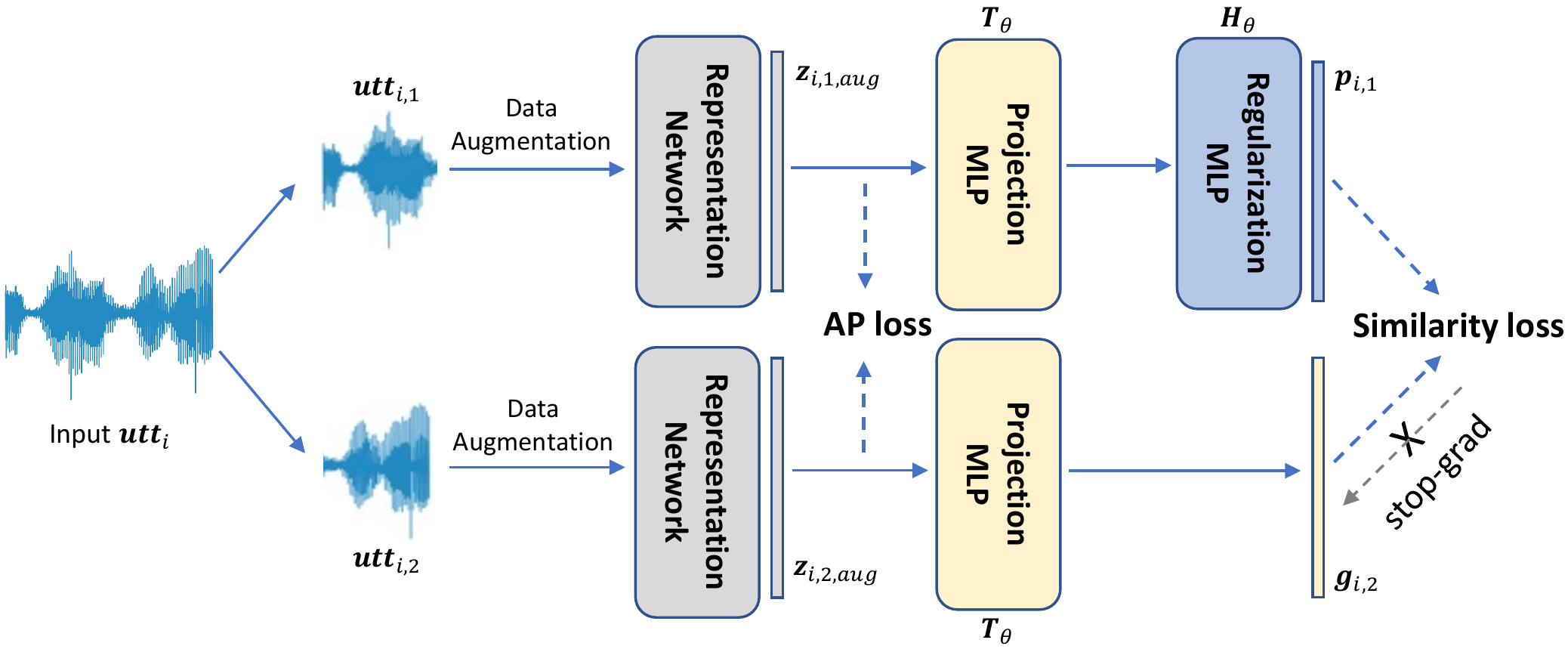}
}
\vspace{-2.0mm}
\caption{Overview of the proposed self-supervised speaker representation learning framework.} 
\label{fig:system}
\end{figure*}

\vspace{-1.0ex}
\section{Related work}
\vspace{-1.0ex}
As a subset of unsupervised learning, self-supervised learning has been explored in the computer vision domain. Recently, contrastive learning methods gained attention and greatly diminished the gap between the performance of supervised learning and self-supervised learning. Most of them are based on the InfoNCE loss with the core idea of attracting the positive sample pairs and spreading the negative samples pairs. 
\hq{However, a large amount of negative samples are usually required to achieve expected performance. }
However, they usually require plenty of negative pairs to work well. For example, MoCo introduces and maintains a dynamic queue with large size to store more negative samples. SimCLR directly utilizes a large batch size to provide sufficient negative samples from the current batch. Accordingly, these methods result in a large overhead in memory requirements.
\hq{why does the large neg samples is important?}

\hq{in speaker V domain, ** has been approved to show **}
For the speaker verification task, it is reasonable to apply contrastive learning since its loss function encourages the learned speaker embeddings to be close for positive pairs and distant for negative pairs. With this optimization goal, learned embeddings tend to be compact for intra-class variations and separable for inter-class variations. In~\cite{inoue2020semi, huh2020augmentation}, researchers have shown promising results of contrastive learning with their proposed semi-supervised and self-supervised learning frameworks. Moreover, an augmentation adversarial training strategy was proposed in~\cite{huh2020augmentation} to encourage the learned speaker embeddings to be channel-invariant. However, adversarial training with gradient reversal layer is substantially complex and unstable for training. Recently, MoCo and SimCLR were also explored on speaker verification task~\cite{ding2020learning, xia2021self} as self-supervised, semi-supervised or pre-training frameworks. In particular, a prototypical memory bank was introduced in~\cite{xia2021self} to improve the self-supervised learning performance by adding an intermediate clustering step. However, these frameworks still deeply rely on a great number of negative samples to work well. Furthermore, researchers in~\cite{nagrani2020disentangled, chung2020seeing} explored cross-modal self-supervised learning methods to learn discriminative speaker embeddings. Different from these methods, this work aims to explore a novel approach without involving the above complex training strategies and architectures. The proposed regularization strategy can use only positive pairs to substantially boost the performance of self-supervised learning with fewer sample pairs and simpler architecture.

\vspace{-1.0ex}
\section{Methodology}
\vspace{-1.0ex}
\hq{too randdom}
In this section, we introduce our proposed approach for self-supervised speaker representation learning. Firstly, to obtain good speaker embeddings, we should ensure that the learned speaker embeddings have the characteristic of intra-class compactness and inter-class separation. Thus, we introduce a self-supervised learning module to encourage latent representations of positive pairs to be close and those of negative pairs to be separate. Meanwhile, to regularize the self-supervised learning module, we propose SSReg to learn robust and speaker-discriminative embeddings by using positive pairs only. Collaborating with the proposed online data augmentation strategy, our approach can significantly improve the performance and consistency of the whole self-supervised speaker verification system. 

%
%

\vspace{-2.0ex}
\subsection{Self-supervised learning framework}
\vspace{-1.0ex}
The overview of our framework is illustrated in Fig. \ref{fig:system}. It consists of the self-supervised learning module and the regularization module. The self-supervised learning module is comprised of (i) a representation network and (ii) the self-supervised angular prototypical loss (AP); the self-supervised regularization module contains (iii) a projection MLP, (iv) a regularization MLP, and (v) the self-supervised regularization loss. The representation network and projection MLP are shared between the upper and lower branches. The representation network is responsible for extracting speaker embeddings, the projection MLP provides non-linear transformations for embeddings. Assuming for each mini-batch, we randomly select $M$ utterances without label information, denoted as $utt_1$, $utt_2$......$utt_M$. For each utterance $utt_i$, we randomly segment it as two non-overlapping segments and apply random data augmentation on them, denoted as $utt_{i,1,aug}$ and $utt_{i,2,aug}$ $(1\leq i \leq M)$. Random data augmentation plays an important role which can provide various channel characteristics to the input and prevent the network from learning the similarity of channel information between segments instead of speaker characteristics. With the data augmentation, extracted speaker embeddings of the first and second segments from utterances of the $i$-th speaker can be represented as $\mathbf{z}_{i,j,aug}$ ($j$=1 or 2). Self-supervised AP loss~\cite{chung2020defence} is applied to maximize the similarity between those embeddings from the same utterance (positive pairs) and minimize the similarity between embeddings from different utterances (negative pairs). The distance metric and loss function are defined as below,

\vspace{-1.0ex}
\begin{equation}
\begin{aligned}
S(\mathbf{z}_{i}, \mathbf{z}_{j})={w \times cosine(\mathbf{z}_{i}, \mathbf{z}_{j})+b} \end{aligned}
\end{equation}

\vspace{-2.5ex}
\begin{equation}
\begin{aligned}
L_{ssl\_ap} =-\frac{1}{N} \sum_{i=1}^{N} \log \frac{\exp \left(S\left(\mathbf{z}_{i, 1,aug}, \mathbf{z}_{i, 2,aug}\right)\right)}{\sum_{j=1}^{N} \exp \left(S\left(\mathbf{z}_{i, 1,aug}, \mathbf{z}_{j,2,aug}\right)\right)}\label{con:AP}   
\end{aligned}
\end{equation}
where $w$ and $b$ denote learnable weight and bias, respectively. $N$ in Equation \ref{con:AP} represents the number of speakers.
\vspace{-1mm}

\vspace{-1.0ex}
\subsection{DNN-based regularization for self-supervised learning}

\vspace{-1.0ex}
\hq{The contrastive module with the AP loss extracts the speaker embedding without extra labeling work. We assume segments from the same utterance are from the same speaker, while segments from different utterances have different identities. However, without labeling information, during training, it is possible that pairs of input segments within a mini-batch are from same speakers. This can be regard as a source of noise during training and contributes to the errors of obtaining a robust speaker embedding. We name this as "noisy labeling error". This error can be alleviated by using a smaller batch size during network training, however, the efficiency of learning will unavoidably degrade with fewer negative samples.

To solve this issue, we add ***, }


The self-supervised learning module with the AP loss extracts speaker embeddings without extra labeling work. Data augmentation can enhance the robustness of speaker embedding networks, since it enables segments to contain different extrinsic variabilities. However, channel information and other extrinsic factors are often remained in speaker embeddings and they are hard to be removed completely. Therefore, we explore and propose a novel regularization strategy to alleviate the interference of channel information and improve the robustness of the extracted speaker embeddings for self-supervised speaker representation learning. In particular, the proposed SSReg only focuses on positive pairs without using negative pairs during self-supervised training.

As shown in Fig. \ref{fig:system}, \hq{can you show those annotations in the figure?}speaker embeddings are processed by the following projection MLP $T_{\theta}$ and the regularization MLP $H_{\theta}$ in the next stage. Two output vectors of upper and lower branches are denoted as $p_{i,1}$ $\triangleq$ $H_{\theta}$ $(T_\theta(\mathbf{z}_{i,1,aug}))$ and $g_{i,2}$ $\triangleq$ $(T_\theta(\mathbf{z}_{i,2,aug}))$ for 
a positive pair. The shared projection MLP maps embeddings to the latent space by a non-linear transformation, and the regularization MLP in the upper branch will learn to reconstruct the transformed embedding $g_{i,2}$ from $p_{i,1}$. Ideally, speaker embeddings should contain speaker information only without other speaker-unrelated extrinsic factors, such as channel information. Thus, good speaker embeddings extracted from the same speaker should be very similar or even identical, regardless of their speech contents and channel environments. Towards this direction, we propose the DNN-based regularization strategy to regularize the self-supervised model training. To promote the invariance of embeddings from the same speaker, we hope that one speaker embedding can be reconstructed and matched by the other one extracted from the different non-overlapping segment of the same utterance, even after a non-linear transformation. Thus, we minimize the negative similarity between latent representations of positive pairs. We define the regularization loss function as,

\vspace{-0.5ex}
\begin{equation}
\begin{aligned}
D(p_{i,1},g_{i,2})=-\frac{p_{i,1}}{\left\|p_{i,1}\right\|_{2}} \cdot \frac{g_{i,2}}{\left\|g_{i,2}\right\|_{2}}\label{con:Sim}
\end{aligned}
\end{equation}
where $\left \| \cdot \right \|_{2}$ denotes $l_2$ norm. Actually, this cosine similarity form is equivalent to the mean square error between two $l_2$ normalized vectors, up to a scale of 2. The loss function encourages the reconstructed latent representation of one augmented segment as similar as the other one within a positive pair. After data augmentation, they are entirely different in channel environments, but from the same speaker. Consequently, only speaker-related information in latent representations contributes to the similarity between segments within positive pairs. When maximizing the similarity during training, latent representations will learn to remain more speaker-related information to meet this optimization goal. In this way, speaker-unrelated information can be removed and the learned speaker embeddings are more robust to extrinsic variabilities. Furthermore, we consider a symmetrized loss for Equation \ref{con:Sim} by changing the order of input segments:

\vspace{-3.0ex}
\begin{equation}
\begin{aligned}
L_{SSReg}=\frac{1}{M} \sum_{i=1}^{M}(\frac12D(p_{i,1},g_{i,2})+\frac12D(p_{i,2},g_{i,1}))
\end{aligned}
\end{equation}
The loss function computes the similarity for every positive pair twice with opposite input order and it is averaged over all input utterances. To avoid mode collapse when using positive pairs only, we introduce the stop-gradient (sg) operation which is applied on the lower branch as demonstrated in Fig. \ref{fig:system}. The symmetrized loss for regularization can be modified as,

\vspace{-3.0ex}
\begin{equation}
\begin{aligned}
L_{SSReg}=\frac{1}{M} \sum_{i=1}^{M}(\frac12D(p_{i,1},sg(g_{i,2}))+\frac12D(p_{i,2},sg(g_{i,1})))
\end{aligned}
\end{equation}
which means that we treat $g_{i,2}$ and $g_{i,1}$ as constants in the first and the second term of the regularization loss respectively.   

Therefore, the overall objective function is defined as a weighed sum of these two loss functions:

\vspace{-2.0ex}
\begin{equation}
\begin{aligned}
L=L_{ssl\_ap}+\lambda L_{SSReg}
\end{aligned}
\end{equation}
where $\lambda$ is the weight for SSReg.

\vspace{-1.0ex}
\section{Experimental setting}
\vspace{-1.5ex}
\subsection{Datasets and feature extraction}
\vspace{-1.0ex}
We train our regularized self-supervised speaker verification system on the development set of VoxCeleb2~\cite{chung2018voxceleb2}, which consists of around 1 million utterances from 5994 speakers. Performance of all systems are evaluated on the test set of VoxCeleb1~\cite{nagrani2017voxceleb}. Additionally, we also study the impacts of different online data augmentation strategies on our self-supervised speaker verification system. To investigate the effectiveness of data augmentation, models are trained with different data augmentation strategies on the development set of VoxCeleb1 which contains 148,642 utterances for 1211 speakers. No speaker labels are used during training in all experiments. 

For all the systems, we compute 40-dimensional log Mel-filterbanks with a frame-length of 25 ms and 10 ms shift. We apply mean and variance normalization (MVN) to the input. VAD is not applied since training data in the VoxCeleb mostly consists of continuous speech. For each input utterance, two segments of 1.95 seconds are randomly selected without overlapping. 

\vspace{-2.0ex}
\subsection{Data Augmentation}
\vspace{-1.0ex}
Data augmentation has been proven to be crucial for both supervised and self-supervised representation learning. The increasing amount and variability of training data can improve the robustness of a speaker embedding system, and the learned speaker embeddings will be more robust to various extrinsic factors. Therefore, we explore the impact of online data augmentation on our regularized self-supervised learning framework. Reverberation, additive noise, speed perturbation, and SpecAug~\cite{park2019specaugment} are considered collectively in our online augmentation strategy. In conventional speaker recognition systems, off-line data augmentation is commonly used to provide only one kind of augmentation to a single input utterance and all augmented data are stored in advance for training. Instead, our online data augmentation allows multiple different augmentations randomly and collectively applied to each input utterance. We use MUSAN corpus~\cite{snyder2015musan} with SNR between 3 to 15 for additive noise and RIR (room impulse response)~\cite{ko2017study} for reverberation.

In the experiments, we explore the effectiveness of following data augmentation strategies, including (1) no data augmentation, (2) reverberation only, (3) additive noise only, (4) reverberation + additive noise, (5) reverberation + additive noise + SpecAug, and (6) reverberation + additive noise + speed perturbation + SpecAug.   

\vspace{-2.0ex}
\subsection{Model configurations and implementation details}
\vspace{-1.0ex}
We exploit the Thin-ResNet34~\cite{chung2020defence} with self-attentive pooling (SAP)~\cite{cai2018exploring} as the representation network, followed by a 512-d fully connected (FC) layer. The projection MLP consists of two FC layers with hidden size of 512. Batch normalization (BN) and ReLU are applied after the first FC layer, and only BN is performed for the second FC layer. The regularization MLP contains two FC layers as well, with BN and ReLU performed on the first FC layer. Inspired by SimSiam~\cite{chen2021exploring}, we use a bottleneck structure for the regularization MLP. The dimension of its input and output is 512, and its hidden layer’s dimension is set to 128. In this way, it can behave like an auto-encoder and might encourage the regularization MLP to digest the information. 

For model training, we use the SGD optimizer with an initial learning rate of 3e-3 and momentum of 0.9. The learning rate is reduced to 4e-5 with a cosine learning rate decay scheduler. The size of extracted speaker embeddings is 512 and the batch size is 250 for each GPU. We use two evaluation metrics to report the system performance: Equal Error Rate (EER) and minimum Detection Cost Function (minDCF) with $p_{target}$= 0.05. Cosine similarity scoring is used to evaluate the performance in the testing phase.

\vspace{-1.0ex}
\section{Experimental results and discussions}
\subsection{Effectiveness of data augmentation}
In the experiments, we investigate the impacts of data augmentation strategies on our self-supervised learning approach. We compare the performance of our framework applied with alternate data augmentations. From table 1, we can observe that adding reverberation and additive noise independently achieve 15.23\% and 14.63\% EER, respectively. Then, randomly applying reverberation and additive noise simultaneously can considerably boost the performance to 9.22\% EER, with a relative 60.3\% improvement in EER compared to the no augmentation applied system. With this data augmentation strategy, our system trained on the VoxCeleb1 dev set can achieve competitive performance and even outperform some previous methods which are trained on the VoxCeleb2. It indicates the effectiveness of our data augmentation strategy and the remarkable performance of our regularized self-supervised learning framework. We also observe that adding SpecAug and speed perturbation cannot consistently gain performance, which is different from our empirical observation for supervised speaker verification. Our results indicate that a suitable and stronger data augmentation strategy is crucial to the self-supervised speaker representation learning framework. It dynamically provides various extrinsic variabilities during training to help reduce the interference of speaker-unrelated factors and improve the robustness of speaker embedding networks.  

\vspace{-2mm}
\begin{table}[h]
\caption{The performance of self-supervised speaker representation learning with different data augmentation strategies. R: Reverberation, N: Additive Noise.}
\vspace{-2.5mm}
\setlength{\tabcolsep}{1.2mm}{
\renewcommand\arraystretch{1.1}
\scalebox{0.99}{
\begin{tabular}{cccc}
\hline \textbf{Augmentation} & \textbf{Corpus} & \textbf{EER(\%)} & \textbf{minDCF} \\
\hline AP {~\cite{huh2020augmentation}} & \multirow{2}{*} { Vox2 } & $9.56$ & $0.511$ $(\mathrm{p}$=$0.05)$ \\
AP + Ch {~\cite{zhang2021contrastive}} & & ${9.23}$ & ${0.6 4 6}$ $(\mathrm{p}$=$0.01)$\\
\hline AP + Ch {~\cite{zhang2021contrastive}} & {Vox1} & $11.07$ & $0.700$ $(\mathrm{p}$=$0.01)$ \\
\hline
No Aug & & $23.21$ & $0.770$ \\
R & & $15.23$ & $0.666$ \\
N & \multirow{2}{*} {Vox1} & $14.63$ & $0.648$ \\
R + N & & $\mathbf{9.22}$ & $\mathbf{0.483}$ \\
R + N + SpecAug & & $9.66$ & $0.500$ \\
R + N + SpecAug + Speed & & $9.98$ & $0.513$ \\
\hline
\end{tabular}}
}
\vspace{-3mm}
\end{table}

\vspace{-1.1ex}
\subsection{Evaluation of speaker embeddings}
\vspace{-0.1ex}
We investigate the performance of our regularized self-supervised learning framework and evaluate it on the VoxCeleb1 test set. We compare our method to~\cite{inoue2020semi, nagrani2020disentangled, chung2020seeing, huh2020augmentation, xia2021self, zhang2021contrastive} which are some recently proposed unsupervised, self-supervised learning architectures and frameworks for speaker representation learning. As shown in Table 2, applying the proposed SSReg independently without self-supervised AP loss can achieve 14.50\% EER, which outperforms some previous works (e.g., Disent, CDDL, GCL, i-vector, and SimCLR). Different from contrastive learning, SSReg only considers the similarity between positive pairs and it does not rely on negative pairs.  
The results suggest that without utilizing any negative samples, our approach shows its potential for learning meaningful speaker embeddings in a self-supervised way. By incorporating SSReg with the self-supervised learning module, our method can generate superior performance and outperform previous works by achieving 6.99\% EER. Compared to AP+AAT and AP+Ch, SSReg can notably improve the performance with relative 19.2\% and 15.6\% reduction in EER, respectively. Without a momentum encoder, a huge number of negative samples (10000 used in~\cite{xia2021self}), or an intermediate clustering step, our method still outperforms MoCo+WavAug and MoCo+WavAug (ProtoNCE) with relative 19.0\% and 15.1\% reduction in EER, respectively. Therefore, the results demonstrate that our method is more effective and robust for self-supervised speaker verification even with a simpler architecture.

\begin{table}[t]
\caption{The performance of self-supervised speaker representation learning evaluated on the Vox1 test set using our method.}
\vspace{-2.0mm}
\setlength{\tabcolsep}{1.6mm}{
\renewcommand\arraystretch{1.05}
\scalebox{0.97}{
\begin{tabular}{l|cc}
\hline \textbf{Systems} & \textbf{EER(\%)} & \textbf{minDCF} \\
\hline Disent {~\cite{nagrani2020disentangled}} & $22.09$ & $-$ \\
CDDL {~\cite{chung2020seeing}} & $17.52$ & $-$ \\
GCL {~\cite{inoue2020semi}} & $15.26$ & $-$ \\
i-vector {~\cite{huh2020augmentation}} & $15.28$ & $0.627$ $(\mathrm{p}$=$0.05)$ \\
AP {~\cite{huh2020augmentation}} & $9.56$ & $0.511$ $(\mathrm{p}$=$0.05)$ \\
AP + AAT {~\cite{huh2020augmentation}} & $8.65$ & $0.454$ $(\mathrm{p}$=$0.05)$ \\
AP + Ch {~\cite{zhang2021contrastive}} & $8.28$ & $0.610$ $(\mathrm{p}$=$0.01)$ \\
SimCLR {~\cite{xia2021self}} & $18.14$ & $0.810$ $(\mathrm{p}$=$0.01)$ \\
MoCo + WavAug {~\cite{xia2021self}} & $8.63$ & $0.640$ $(\mathrm{p}$=$0.01)$ \\
MoCo + WavAug (ProtoNCE) {~\cite{xia2021self}} & $8.23$ & $0.590$ $(\mathrm{p}$=$0.01)$ \\
\hline Ours (w/o AP) & $14.50$ & $0.621$ \\
Ours (Full model) & $\mathbf{6.99}$ & $\mathbf{0.434}$ \\
\hline
\end{tabular}}
}
\vspace{-3mm}
\end{table}



\vspace{-1ex}
\begin{table}[h]
\caption{The effect of the weight $\lambda$ for regularization on self-supervised speaker verification performance.}
\vspace{-5.5mm}
\setlength{\tabcolsep}{4mm}{
\begin{center}
\renewcommand\arraystretch{1.0}

\begin{tabular}{l|cc}
\hline \textbf{Weight} $\lambda$ & \textbf{EER} (\%) & \textbf{minDCF} \\
\hline \makecell{$0$} & $9.13$ & $0.490$ \\
\makecell{$0.01$} & $7.10$ & $0.448$ \\
\makecell{$0.08$} & $\mathbf{6.99}$ & ${0.434}$ \\
\makecell{$0.1$} & $7.11$ & $0.445$ \\
\makecell{$0.3$} & $7.22$ & $0.441$ \\
\makecell{$0.5$} & ${7.30}$ & $\mathbf{0.432}$ \\
\hline
\end{tabular}
\end{center}}

\vspace{-3mm}
\end{table}

\vspace{-2mm}
Moreover, we also conduct experiments to investigate the impact of SSReg on self-supervised speaker verification performance. From Table 3, we observe that applying SSReg can greatly improve the performance. It achieves the best performance with 6.99\% EER at $\lambda$=0.08. There is a relative 23.4\% reduction in EER compared to the system without applying regularization. The results confirm the effectiveness of the proposed method in speaker verification task. 

\vspace{-0.8ex}
\section{Conclusions}
\vspace{-1.0ex}
In this paper, we propose a self-supervised learning framework that consists of the self-supervised learning module and the self-supervised regularization module. With our well-designed online data augmentation strategy, the proposed self-supervised learning framework can effectively leverage unlabeled data to learn robust and speaker-discriminative embeddings. In particular, the proposed DNN-based self-supervised regularization enables a simple Siamese network to learn robust speaker embeddings by using positive pairs only. 
Moreover, SSReg can significantly improve the performance of self-supervised learning. Experimental results demonstrate that our method outperforms previous works with a simpler but more effective framework. This work can provide a new direction to study the speaker verification task with self-supervised learning methods.

\vfill\pagebreak


\bibliographystyle{IEEEbib}
\bibliography{icassp21}

\begin{thebibliography}{10}

\bibitem{snyder2017deep}
David Snyder, Daniel Garcia-Romero, Daniel Povey, and Sanjeev Khudanpur,
\newblock ``Deep neural network embeddings for text-independent speaker
  verification.,''
\newblock in {\em Interspeech}, 2017, pp. 999--1003.

\bibitem{snyder2018x}
David Snyder, Daniel Garcia-Romero, Gregory Sell, Daniel Povey, and Sanjeev
  Khudanpur,
\newblock ``X-vectors: Robust dnn embeddings for speaker recognition,''
\newblock in {\em ICASSP}, 2018, pp. 5329--5333.

\bibitem{desplanques2020ecapa}
Brecht Desplanques, Jenthe Thienpondt, and Kris Demuynck,
\newblock ``Ecapa-tdnn: Emphasized channel attention, propagation and
  aggregation in tdnn based speaker verification,''
\newblock {\em arXiv preprint arXiv:2005.07143}, 2020.

\bibitem{wan2018generalized}
Li~Wan, Quan Wang, Alan Papir, and Ignacio~Lopez Moreno,
\newblock ``Generalized end-to-end loss for speaker verification,''
\newblock in {\em ICASSP}, 2018, pp. 4879--4883.

\bibitem{chung2020defence}
Joon~Son Chung, Jaesung Huh, Seongkyu Mun, Minjae Lee, Hee~Soo Heo, Soyeon
  Choe, Chiheon Ham, Sunghwan Jung, Bong-Jin Lee, and Icksang Han,
\newblock ``In defence of metric learning for speaker recognition,''
\newblock {\em arXiv preprint arXiv:2003.11982}, 2020.

\bibitem{wang2018cosface}
Hao Wang, Yitong Wang, Zheng Zhou, Xing Ji, Dihong Gong, Jingchao Zhou, Zhifeng
  Li, and Wei Liu,
\newblock ``Cosface: Large margin cosine loss for deep face recognition,''
\newblock in {\em Proceedings of the IEEE conference on computer vision and
  pattern recognition}, 2018, pp. 5265--5274.

\bibitem{deng2019arcface}
Jiankang Deng, Jia Guo, Niannan Xue, and Stefanos Zafeiriou,
\newblock ``Arcface: Additive angular margin loss for deep face recognition,''
\newblock in {\em Proceedings of the IEEE/CVF Conference on Computer Vision and
  Pattern Recognition}, 2019, pp. 4690--4699.

\bibitem{cai2018exploring}
Weicheng Cai, Jinkun Chen, and Ming Li,
\newblock ``Exploring the encoding layer and loss function in end-to-end
  speaker and language recognition system,''
\newblock in {\em Proc. Odyssey 2018 The Speaker and Language Recognition
  Workshop}, 2018, pp. 74--81.

\bibitem{okabe2018attentive}
Koji Okabe, Takafumi Koshinaka, and Koichi Shinoda,
\newblock ``Attentive statistics pooling for deep speaker embedding,''
\newblock {\em arXiv preprint arXiv:1803.10963}, 2018.

\bibitem{sang2020open}
Mufan Sang, Wei Xia, and John~HL Hansen,
\newblock ``Open-set short utterance forensic speaker verification using
  teacher-student network with explicit inductive bias,''
\newblock {\em Proc. Interspeech 2020}, pp. 2262--2266, 2020.

\bibitem{bhattacharya2019generative}
Gautam Bhattacharya, Joao Monteiro, Jahangir Alam, and Patrick Kenny,
\newblock ``Generative adversarial speaker embedding networks for domain robust
  end-to-end speaker verification,''
\newblock in {\em ICASSP}. IEEE, 2019, pp. 6226--6230.

\bibitem{sang2021deaan}
Mufan Sang, Wei Xia, and John~HL Hansen,
\newblock ``Deaan: Disentangled embedding and adversarial adaptation network
  for robust speaker representation learning,''
\newblock in {\em ICASSP}. IEEE, 2021, pp. 6169--6173.

\bibitem{chen2020simple}
Ting Chen, Simon Kornblith, Mohammad Norouzi, and Geoffrey Hinton,
\newblock ``A simple framework for contrastive learning of visual
  representations,''
\newblock in {\em International conference on machine learning}. PMLR, 2020,
  pp. 1597--1607.

\bibitem{he2020momentum}
Kaiming He, Haoqi Fan, Yuxin Wu, Saining Xie, and Ross Girshick,
\newblock ``Momentum contrast for unsupervised visual representation
  learning,''
\newblock in {\em Proceedings of the IEEE/CVF Conference on Computer Vision and
  Pattern Recognition}, 2020, pp. 9729--9738.

\bibitem{grill2020bootstrap}
Jean-Bastien Grill, Florian Strub, Florent Altch{\'e}, Corentin Tallec,
  Pierre~H Richemond, Elena Buchatskaya, Carl Doersch, Bernardo~Avila Pires,
  Zhaohan~Daniel Guo, Mohammad~Gheshlaghi Azar, et~al.,
\newblock ``Bootstrap your own latent: A new approach to self-supervised
  learning,''
\newblock {\em arXiv preprint arXiv:2006.07733}, 2020.

\bibitem{chen2021exploring}
Xinlei Chen and Kaiming He,
\newblock ``Exploring simple siamese representation learning,''
\newblock in {\em Proceedings of the IEEE/CVF Conference on Computer Vision and
  Pattern Recognition}, 2021, pp. 15750--15758.

\bibitem{inoue2020semi}
Nakamasa Inoue and Keita Goto,
\newblock ``Semi-supervised contrastive learning with generalized contrastive
  loss and its application to speaker recognition,''
\newblock in {\em 2020 Asia-Pacific Signal and Information Processing
  Association Annual Summit and Conference (APSIPA ASC)}. IEEE, 2020, pp.
  1641--1646.

\bibitem{huh2020augmentation}
Jaesung Huh, Hee~Soo Heo, Jingu Kang, Shinji Watanabe, and Joon~Son Chung,
\newblock ``Augmentation adversarial training for unsupervised speaker
  recognition,''
\newblock in {\em Workshop on Self-Supervised Learning for Speech and Audio
  Processing, NeurIPS}, 2020.

\bibitem{nagrani2020disentangled}
Arsha Nagrani, Joon~Son Chung, Samuel Albanie, and Andrew Zisserman,
\newblock ``Disentangled speech embeddings using cross-modal
  self-supervision,''
\newblock in {\em ICASSP}. IEEE, 2020, pp. 6829--6833.

\bibitem{chung2020seeing}
Soo-Whan Chung, Hong~Goo Kang, and Joon~Son Chung,
\newblock ``Seeing voices and hearing voices: learning discriminative
  embeddings using cross-modal self-supervision,''
\newblock {\em arXiv preprint arXiv:2004.14326}, 2020.

\bibitem{xia2021self}
Wei Xia, Chunlei Zhang, Chao Weng, Meng Yu, and Dong Yu,
\newblock ``Self-supervised text-independent speaker verification using
  prototypical momentum contrastive learning,''
\newblock in {\em ICASSP}. IEEE, 2021, pp. 6723--6727.

\bibitem{ding2020learning}
Ke~Ding, Xuanji He, and Guanglu Wan,
\newblock ``Learning speaker embedding with momentum contrast,''
\newblock {\em arXiv preprint arXiv:2001.01986}, 2020.

\bibitem{chung2018voxceleb2}
Joon~Son Chung, Arsha Nagrani, and Andrew Zisserman,
\newblock ``Voxceleb2: Deep speaker recognition,''
\newblock {\em Proc. Interspeech 2018}, pp. 1086--1090, 2018.

\bibitem{nagrani2017voxceleb}
Arsha Nagrani, Joon~Son Chung, and Andrew Zisserman,
\newblock ``Voxceleb: A large-scale speaker identification dataset,''
\newblock {\em Proc. Interspeech 2017}, pp. 2616--2620, 2017.

\bibitem{park2019specaugment}
Daniel~S Park, William Chan, Yu~Zhang, Chung-Cheng Chiu, Barret Zoph, Ekin~D
  Cubuk, and Quoc~V Le,
\newblock ``Specaugment: A simple data augmentation method for automatic speech
  recognition,''
\newblock {\em arXiv preprint arXiv:1904.08779}, 2019.

\bibitem{snyder2015musan}
David Snyder, Guoguo Chen, and Daniel Povey,
\newblock ``Musan: A music, speech, and noise corpus,''
\newblock {\em arXiv preprint arXiv:1510.08484}, 2015.

\bibitem{ko2017study}
Tom Ko, Vijayaditya Peddinti, Daniel Povey, Michael~L Seltzer, and Sanjeev
  Khudanpur,
\newblock ``A study on data augmentation of reverberant speech for robust
  speech recognition,''
\newblock in {\em ICASSP}. IEEE, 2017, pp. 5220--5224.

\bibitem{zhang2021contrastive}
Haoran Zhang, Yuexian Zou, and Helin Wang,
\newblock ``Contrastive self-supervised learning for text-independent speaker
  verification,''
\newblock in {\em ICASSP}. IEEE, 2021, pp. 6713--6717.

\end{thebibliography}


\end{document}